\documentstyle[sprocl,epsf]{article}

\def\D	{{\bf D}}
\def\B	{{\bf B}}
\def\E	{{\bf E}}
\def\A	{{\bf A}}
\def\x	{{\bf x}}
\def\CA{C_{\rm A}}
\def\FAC{{\hbox{\tiny FAC}}}
\def\NLLO{{\hbox{\tiny NLLO}}}
\def\Const{{\cal C}}

\def\alphaw{\alpha_{\rm w}}

\def\bzeta{\mbox{\boldmath$\zeta$}}

\def\bigdrangle{\bigr\rangle\!\bigr\rangle}
\def\bigdlangle{\bigl\langle\!\bigl\langle}

\begin{document}

\title{Non-perturbative dynamics of hot non-Abelian gauge fields$^*$}

\footnotetext
    {%
    $^*$ Based on talks presented at
    ``Quarks-2000'', Pushkin, Russia, May 2000, and
    ``Strong and Electroweak Matter 2000'',
    Marseilles, France, June 2000.
    This summary describes work performed in collaboration
    with Peter Arnold.
    A more detailed account of this material may be found
    in references~1 and 2.
    }

\author{Laurence G. Yaffe}

\address
    {%
    Department of Physics,
    University of Washington,
    Seattle, WA 98195-1560, USA
    \\E-mail: yaffe@phys.washington.edu
    }

%

\maketitle

\abstracts
    {%
    The dynamics of high temperature gauge fields,
    on scales relevant for non-perturbative phenomena
    such as electroweak baryogenesis,
    may be described by a remarkably simple effective theory.
    This theory, which takes the form of a local, stochastic,
    classical Yang-Mills theory, depends on a single parameter,
    the non-Abelian (or ``color'') conductivity.
    This effective theory has recently been shown to be valid
    to next-to-leading-log order (NLLO), provided one uses an
    improved NLLO value for the non-Abelian conductivity.
    Comparisons of numerical simulations using this NLLO effective
    theory and a more microscopic effective theory agree
    surprisingly well.
    }

In high temperature non-Abelian gauge theories,
non-perturbative properties
such as the rate of baryon number violation or the
friction on an expanding bubble wall
are sensitive to the dynamics of low frequency,
long wavelength gauge field fluctuations.\cite {alpha5,bubble}
The relevant spatial scale for non-perturbative fluctuations
has long been understood to be $1/(g^2 T)$ where,
for electroweak applications, $g$ is the $SU(2)_{\rm w}$
gauge coupling.
The corresponding time scale (for weakly coupled theories)
is parametrically longer, and turns out to be $1/(g^4 T \ln g^{-1})$.%
\cite {alpha5,bodeker,Blog1}

The dynamics of low-frequency gauge field fluctuations
is approximately described by a remarkably simple
effective theory, whose equation of motion is
\begin {equation}
    \D \times \B = \sigma \, \E - \bzeta \,.
\label {eff1}
\end {equation}
Here, $\bzeta$ is Gaussian noise, {\em i.e.},
a random (adjoint valued) vector field, with variance
\begin {equation}
   \bigdlangle \zeta_i(t,\x) \, \zeta_j(t',\x') \bigdrangle
    =
    2 \sigma T \, \delta_{ij} \, \delta(t{-}t') \, \delta^{(3)}(\x{-}\x') .
\label {noise}
\end {equation}
Eqs.~(\ref {eff1}) and (\ref {noise}) define stochastic 3-dimensional
gauge theory; it has a long history of study for reasons having
nothing to do with high temperature dynamics.\cite {stochastic}
The parameter $\sigma$ is naturally termed a
non-Abelian (or ``color'') conductivity.
In $A_0 = 0$ gauge, Eq.~(\ref {eff1}) becomes a first-order
stochastic partial differential equation,
$
    -\sigma \, (d\A / dt) = \D \times \B + \bzeta
$,
and may be regarded as a natural generalization of the
equation of motion of an over-damped oscillator in the presence
of thermal noise, with $\sigma$ playing the role of the damping
constant.

This effective theory was first derived
by D.~B\"odeker,\cite {bodeker}%
\footnote {For other derivations,
see also Refs.~\cite {Blog1,moore,manuel,basagoiti,b&i}.}
starting from the underlying high temperature quantum field theory
and carefully integrating out the effects of thermal (and quantum)
fluctuations on shorter scales,
assuming that the gauge coupling
$g$ is so small that corrections suppressed by powers of $g$,
or even powers of $1/\ln g^{-1}$, are negligible.
In other words, the effective theory was the result of
a {\em leading-log} analysis.
Within this approximation, one finds\,\cite {bodeker,Blog1}
that the color conductivity
$
    \sigma = {m^2 / (3\gamma)}
$,
up to relative corrections of order $1/\ln g^{-1}$,
where $m$ is the leading-order Debye screening mass
[equal to $\sqrt {11/6} \> gT$ in minimal electroweak theory],
and $\gamma$ is the hard gauge boson damping rate\,\cite {gammag}
which, to leading-log order, is $\CA \alpha T \ln g^{-1}$.
Using this theory, Moore\,\cite{moore}
has numerically simulated the topological transition
(or baryon violation)
rate for electroweak theory, obtaining
\begin {equation}
   \Gamma \approx (10.8 \pm 0.7)
       \left( {2\pi \over 3\sigma} \right)
       (\alpha T)^5 \,.
\label{eq:GammaB}
\end {equation}

The presence of corrections suppressed just by powers of
$1/\ln g^{-1}$ reflects the fact that the effective theory
defined by Eqs.~(\ref {eff1}) and (\ref {noise}) is only valid
on spatial scales large compared to the inverse damping rate
$\gamma^{-1}$.
As in any effective theory, there will be higher order corrections
suppressed by powers of $(k/\Lambda)$,
where $k$ is the momentum scale of interest,
and $\Lambda$ is the defining scale of the effective theory.
For this application, the scale of interest $k \sim g^2 T$,
while $\Lambda = \gamma \sim g^2 T \ln g^{-1}$.

A natural question to ask is whether one can extend
this leading-log analysis and produce a useful effective
theory that is valid beyond leading-log order.
After all, leading-log results, by themselves, have minimal
practical utility --- there is a huge difference between,
say, $\ln (16\pi/g)$ and $\ln (1/10g)$ for any realistic
value of the gauge coupling.
But this difference is of sub-leading order in $1/\ln g^{-1}$.

The obvious first step is a next-to-leading-log order (NLLO)
treatment, in which all corrections suppressed by one power
of $1/\ln g^{-1}$ are retained, while all higher powers are neglected.
This is what Peter Arnold and I set out to do.

Typically, when one wishes to go to next order in $k/\Lambda$
in any effective theory, one must to do two things:
({\em i\/}) add to the effective theory a set of new higher-dimension operators,
and ({\em ii\/}) re-determine the coefficients of operators in the
effective theory by performing a more accurate matching calculation
in which one compares physical quantities computed in the effective
theory with some more fundamental theory (or experiment).

Fortunately, in this problem a wonderful simplification occurs.
The effects of all higher-dimension corrections to the effective theory
(\ref {eff1}) are suppressed by two or more powers of $1/\ln g^{-1}$.
Ref.~1 describes a simple power counting analysis leading to
this result, as well as a more elaborate (but more thorough) analysis
based on formulating the stochastic effective theory as a supersymmetric
functional integral.
Consequently, the form of the effective theory remains unchanged
to NLLO, one must only calculate its single parameter (namely $\sigma$)
to next-to-leading log order.

The fact that no higher-dimension terms are needed at NLLO
has several other immediate consequences.
Stochastic three-dimensional gauge theory is ultraviolet finite.%
\cite {Blog1,zinnjustin&zwanziger}
Hence, the effective theory (\ref {eff1}) may be discretized
on a lattice and numerically simulated without any
regularization-dependent subtleties or renormalization of parameters.%
\cite {moore}
Because the form of the effective theory remains unchanged at NLLO,
previous numerical simulations of the leading-log theory may
be instantly extended to NLLO accuracy simply by using a suitably
improved value for $\sigma$;
no new numerical simulations are needed.
Because the effective theory remains UV finite at NLLO, this also means
that its one parameter, the color conductivity $\sigma$,
is unambiguously defined at NLLO.
In contrast,
if new higher-dimension terms had been necessary,
this would have destroyed the UV finiteness of the theory, and
$\sigma$ would then become a scheme and scale dependent quantity
(just like quark masses in QCD).
But to NLLO, one may regard $\sigma$ as a well-defined physical quantity.%
\footnote
    {%
    This is significant because, unlike the situation for ordinary
    electrical conductivity, there is no Kubo formula or other
    gauge-invariant physical definition of a non-Abelian conductivity.
    }

The next-to-leading-log determination of the color conductivity
was carried out in Ref.~2.
It requires extending the conventional technology for performing
effective field theory matching calculations to a situation involving
real-time stochastic theories.
And it requires identifying an appropriate physical quantity that may be
evaluated in both the effective and microscopic theories, from which
one may determine $\sigma$ by comparing the two results.
We argued that a suitable Minkowski-space Wilson loop provides an appropriate
physical quantity and showed that operationally, in Coulomb gauge,
this reduces (at NLLO) to matching the low momentum behavior of
the zero frequency gauge field self-energy $\Pi_{00}(0,k)$.

To carry out the matching calculation, we found it convenient
to work with a sequence of three effective theories.
Theory 1 was
a linearized collisionless Boltzmann-Vlasov kinetic theory equivalent
to the usual HTL (``hard-thermal-loop'') effective theory,
valid for momenta and frequencies $k,\omega \ll T$.
Theory 2 was a stochastic, collisional, linearized kinetic theory
valid for $\omega \ll k \ll T$.
And Theory 3 was the final effective theory
given by Eqs.~(\ref {eff1}) and (\ref {noise}) ---
a diffusive Langevin equation, valid for $\omega \ll k \ll \gamma$.
The structure of our result is clearest if one writes an expression
for $\sigma^{-1}$ (the ``color resistivity'') rather than $\sigma$
directly.
We found,
\begin {eqnarray}
   \sigma^{-1}
   &=& {3\CA \alpha T\over m^2}
       \left[\,
       \ln\!\left(m\over\gamma(\mu)\right)
          + \Const
	  + O\!\left(1 \over \ln(1/g)\right)
       \right] ,
\label{eq:final}
\\
\noalign {\hbox {with}}
    \Const &=& 3.0410\cdots \,.
\label {eq:final2}
\end {eqnarray}
Inside the logarithm of (\ref {eq:final}),
$\gamma(\mu)$ is to be understood as the leading-log formula
\begin {equation}
   \gamma(\mu) \approx \CA \alpha T \ln\!\left(m\over\mu\right) ,
\end {equation}
with $\mu$ chosen so that it is of order $\gamma$
(so as to avoid unnecessarily large logarithms in the
description of physics on the scale of $\gamma$).
The residual $\mu$ dependence in the NLLO result
(\ref{eq:final}) only affects that answer at order
$[\ln(m/\gamma)]^{-1} \sim [\ln(1/g)]^{-1}$, which is beyond
the order of this calculation.

To use the NLLO result (\ref {eq:final}) in any practical calculation
one must choose some particular value of $\mu$
and ignore the unknown $O(1/\ln g^{-1})$ corrections.
In the absence of a full next-to-next-to-leading-log analysis,
there is no clearly preferred procedure for determining an ``optimal'' value.
However, one reasonably natural choice is the ``fastest apparent convergence''
(FAC) scheme in which $\mu$ is chosen so that the
next-to-leading order correction vanishes.
This amounts to choosing the scale $\mu_\FAC$ satisfying
$
    \mu_{\FAC} = e^{-\Const} \, \gamma(\mu_{\FAC}) \,.
$
For this choice, the NLLO conductivity is simply
\begin {equation}
   \sigma^{-1}_\NLLO
   = {3\CA \alpha T\over m^2} \,
       \ln\!\left(m\over\mu_{\FAC}\right) \,.
\end {equation}

\begin {figure}[t]
   \begin {center}
      \leavevmode
      \def\epsfsize #1#2{0.85#1}
      \begin {picture}(0,0)
	\put(120,-5){$\gamma/\mu$}
	\put(-45,80){$\displaystyle\ln
                \left({e^\Const m \over \gamma(\mu)}\right)$}
	\put(181,22){\LARGE $\downarrow$}
	\put(181,88){\LARGE $\uparrow$}
      \end {picture}
      \epsfbox {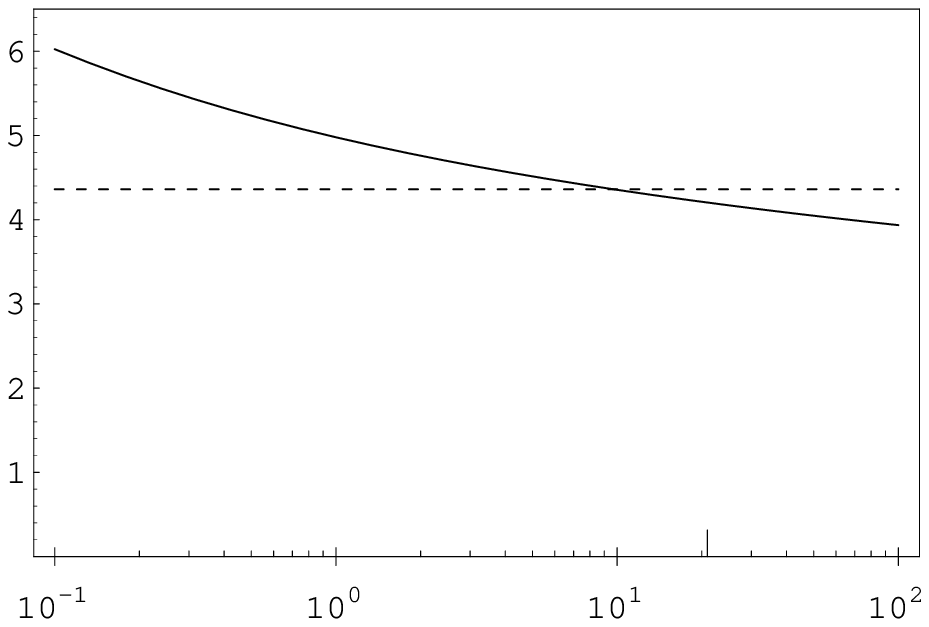}
   \end {center}
   \caption
	{%
	The value of the $\ln \!\left[m / \gamma(\mu)\right]\!{+}\Const$
        factor,
	appearing in the inverse color conductivity
	(\protect\ref {eq:final})
	and in the topological transition rate (\protect\ref {eq:Gam}),
	plotted as a function of $\gamma/\mu$,
	for electroweak theory with a single Higgs doublet and $g^2 = 0.4$.
	The dashed line indicates the value of $4.4$ for which
	the NLLO result for the topological transition rate
	(\protect\ref {eq:Gam}) (with the unknown yet-higher-order
	$1/\ln g^{-1}$ terms neglected) agrees with independently determined
	results from more microscopic numerical simulations.%
	\protect\cite {top-trans,moore-SEWM}
	The arrow on the abscissa indicates the FAC point where
	$\gamma(\mu)/\mu = e^\Const \simeq 20.93$.
	\label {fig}%
	}
\end {figure}

Using our NLLO result (\ref{eq:final}), one may instantly
generalize Moore's numerical result (\ref{eq:GammaB})
for the topological transition rate of hot electroweak theory to NLLO:
\begin {equation}
   \Gamma \approx (10.8 \pm 0.7)
       \left( {g T \over m} \right)^2
       \alphaw^5 \, T^4
       \left[\,
       \ln\!\left(m\over\gamma(\mu)\right)
          + 3.041
	  + O\!\left(1 \over \ln(1/g)\right)
       \right] .
\label {eq:Gam}
\end {equation}

Moore, and others, have also obtained numerical results for the topological
transition rate by using a more microscopic theory (analogous to a lattice
version of what we called Theory 1).\cite {top-trans,moore-SEWM}
Moore also attempted to estimate the size of the NLLO correction to $\Gamma$
by fitting the results of these simulations to the functional form
\begin {equation}
    \Gamma = \kappa \left({gT \over m}\right)^2 \alphaw^5 \, T^4
    \left[\,
       \ln\!\left(m\over g^2 T\right)
          + \delta \,
       \right] ,
\label {eq:guess}
\end {equation}
with the value of $\kappa$ fixed to 10.8, as determined from simulations
of the effective theory.
This led to $\delta \approx 3.6$, with perhaps 20\% uncertainty due to
systematic errors.\cite {moore-SEWM}\,%
\footnote
    {%
    Lattice artifacts exist in these more microscopic simulations
    (due, in part, to the lattice dispersion relation allowing
    unwanted Cherenkov radiation),
    which cause them not to reproduce, precisely, the dynamics
    responsible for NLLO corrections to $\Gamma$.
    These effects have only been crudely estimated, but are
    a major part of the uncertainty in the estimate of $\delta$.
    }
This implies an estimate of $4.4$ for the value of the
$\ln (m / g^2 T) {+} \delta$
factor in (\ref {eq:guess}), which
may be compared with the square bracket appearing in
(\ref {eq:Gam}).
Fig.~\ref{fig} shows this comparison.
The solid line is a plot of
$\ln \!\left[m / \gamma(\mu)\right]\!{+}\Const$ as a function of
$\gamma/\mu$.
The dashed line indicates the value of $4.4$ estimated in
Ref.~\cite {moore-SEWM}.
The arrow on the abscissa indicates the FAC point, where
$\gamma(\mu)/\mu = e^\Const \simeq 20.93$.

The similarity between our NLLO result
and the value
inferred from numerical simulations is remarkable.
There was no obvious {\it a priori\/} reason why it should be a
reasonable approximation to treat logarithms of the gauge coupling
[that is, $\ln(\#/g)$],
as large for physical values of the coupling.
The close agreement between the FAC value of $\mu$ and the precise
point where the curves of Fig.\ \ref{fig}
cross is striking, but probably fortuitous
given the uncertainty in the numerical simulation value.
Nevertheless, it may well be that the characteristic scale for
neglected corrections really is
$e^{-\Const} \gamma$,
and not just $\gamma$ (as one might naively expect).
If true, this would mean that (for electroweak theory)
the expansion in inverse logs actually has a respectably small
expansion parameter of about 0.25.
%

\end{document}